\documentclass[12pt]{article}
 
\usepackage{amssymb}
\usepackage{amsmath}
\usepackage{amsthm}
\usepackage{geometry} 
\usepackage{graphicx}
\usepackage{subcaption}
\usepackage{hyperref}
\usepackage{parskip}
\usepackage{comment}
\usepackage[toc]{appendix}
\usepackage{amsfonts}
\usepackage{xcolor}
\usepackage{chemformula}
\usepackage{lineno}
\usepackage{enumerate}
\theoremstyle{remark}
\newtheorem{remark}{Remark}

\usepackage[
    backend=bibtex,
    style=numeric,
    sorting=nyt,
    maxbibnames=10,
    citestyle=numeric-comp
  ]{biblatex}
\usepackage{authblk}
\begin{document}

\title{A hybrid model of sulphation reactions:   \\stochastic particles   in a random continuum environment}

\author[a]{Nicklas J\"averg\r{a}rd }
\author[b]{Daniela Morale }
\author[b]{Giulia Rui }
\author[a]{Adrian Muntean }
\author[b]{Stefania Ugolini }

\affil[a]{\small  Dept of Mathematics and Computer Science,  University of Karlstad, Sweden\\\texttt{Nicklas.Javergard@kau.se, Adrian.Muntean@kau.se}}
\affil[b]{\small Dept. of Mathematics, University of Milano, Italy\\\texttt{Daniela.Morale@unimi.it,Giulia.Rui@unimi.it, Stefania.Ugolini@unimi.it}}
\maketitle

\begin{abstract} 
 We present a hybrid stochastic-continuum model to study the sulphation of calcium carbonate and the consequent formation of gypsum, a key phenomenon driving marble deterioration.
While calcium carbonate and gypsum are continuous random fields evolving according to random ordinary differential equations, the dynamics of sulfuric acid particles follows Itô-type stochastic differential equations. 
The particle evolution incorporates both strong repulsion between particles via the Lennard-Jones potential, and non-local interactions with the continuum environment. The particle-continuum coupling is also achieved through the chemical reaction, modeled as a Poisson counting process.
We simulate the spatiotemporal evolution of this corrosion process using the Euler-Maruyama algorithm with varying initial data combined with finite elements to take care of the spatial discretization. Despite symmetric initial data, our simulations highlight an uneven progression of corrosion due to the stochastic influences in the model.
\end{abstract}


\section{Introduction}
 The preservation of cultural heritage has gained increasing attention in various scientific communities during recent years, particularly as in highly agglomerated urban areas, marble and other calcium carbonate-based monuments face a quick degradation due to atmospheric pollution. Understanding the mechanisms behind such chemical deterioration is essential for developing effective prevention strategies. Mathematical modeling provides  powerful tools to analyze the involved physico-chemical processes, offering insights that could help predict material degradation. In-depth  investigations of material corrosion using modern mathematical tools raise  as well intriguing theoretical questions. As main guiding theme of our research, a major challenge in forecasting the large-time behavior of degradation processes, as in the case of marble-based monuments, lies in capturing the spatiotemporal variability of real-world conditions while maintaining a balance between complexity and tractability. This motivates us to develop  hybrid mathematical models that have the ability to integrate naturally both stochastic and deterministic modeling components at different scales to better understand and then simulate meaningfully corrosion processes. 
Taking as case-study the durability of marble monuments, the key player in the deterioration process is the sulphation reaction -- sulfur dioxide in the atmosphere is converted in the presence of moist air into sulfuric acid, which then reacts with calcium carbonate (main chemical component in marble), forming gypsum as a byproduct. The accumulation of gypsum results in the formation of black crusts \cite{2024_COMITE},  which are not only aesthetically undesirable, but the occurrence of such crusts may facilitate corrosion. In fact, compared to calcium carbonate, gypsum is significantly more porous, allowing for a deeper penetration of the acid  (and of other unwanted substances). For more chemical point of view details, see, for instance, \cite{2007_Bonazza_Sabbioni,2021_Comite_Fermo_LaRussa,2020_Comite_fermo1,2024_COMITE,Saba_modelsReview}. 
Additionally, external environmental factors such as temperature, relative humidity, and the interaction with metallic pollutants further influence the corrosion rate (and of the corresponding corrosion speed), making the process highly sensitive to local conditions; compare to  \cite{2024_COMITE,Catalyst,FieldExposure} for more context on the matter.

The mathematical modeling of marble sulphation has primarily been studied from a continuum   macroscopic  perspective,  where deterministic models describe the time and space  evolution of the relevant substances by means of balance laws formulated for systems of  partial differential equations \cite{Saba_modelsReview,Natalini_model,Natalini2, Tasnim}. To reflect the intrinsic randomness of pollutant distribution and environmental fluctuations, recent studies have introduced stochastic elements into the modeling framework. In the  very recent works \cite{2024_arxivArceciMoraleUgolini1,Morale_Ugolini_2023,2023_MauMorUgo,MACH2023_AMMU},  the authors couple deterministic reaction models with stochastic boundary conditions in order to capture the inherent variability of the acid particles in the atmosphere and of the microscale interactions at the material-environment interface. Stochastic approaches to chemical reactions have been studied in rather general settings (cf. e.g. \cite{Anderson_ref1,Anderson_ref2}), but in {\em loc. cit.} one does not take into account the spatial dependence required for the interaction with the environment. A fully probabilistic interpretation of the setting described in  \cite{Natalini_model,Natalini2} has been proposed in \cite{2024_arxivMoraleTarquiniUgolini1} relying on  an interacting particle system of the McKean-Vlasov type. Another fully stochastic approach was proposed in \cite{MACH2023_MRU}, where the authors model the dynamics of all molecules involved in the sulphation reaction by stochastic differential equations, with reactions driven by  point process es featuring a state-dependent reaction rate. While these approaches offer a more detailed microscopic perspective, their high complexity makes both analytical and computational studies challenging.
Despite these advancements, challenges remain in balancing the modeling of the intrinsic variability that we experience at the material-environment interface with the complexity of the system and its description, plus in ensuring the flexibility of models to adapt to a wide range of real-world conditions.

In this paper we propose a hybrid modeling approach that couples explicitly the stochastic dynamics at microscale of the acid particles with a continuous description at the macroscale of the time-space evolution of the calcium carbonate and gypsum fields, which are the local environment hosting the degradation process.  Since it is  subject to  the chemical interaction with  acid particles, such continuum evolutions becomes random themselves, as e.g. in \cite{2016_MZCJ,2012_CMF,2019_FL}.  
Specifically, the acid particles dynamics is governed by a system of stochastic differential equations of the It\^o type. The particles  are subject to pairwise interactions via a strongly repulsive potential at close range, and a weaker attraction  at longer distances, and to a non-local interaction with the continuous environment, modeling the different physical properties of calcium carbonate and gypsum. For the pairwise interaction we consider a suitable regularization of the  Lennard-Jones potential   in order to avoid overlapping of particles in a finite time. Indeed, Lennard-Jones potential, even though has the cited good properties,  produces a singular drift and  presents a super linear growing at the origin. On the other hand, the environment gradually changes in time as the densities of calcium carbonate and gypsum evolve according to the deterministic law of mass action.  The coupling with the stochastic particles via their empirical concentration introduces  randomness at the macroscale of the continuum; therefore both calcium and gypsum densities evolve according to random ordinary differential equations. Finally, the lifetime of each acid particle is determined as the  reaction occurrence of  associated Poisson counting processes, whose rate depends locally on the state of the continuum calcium carbonate density  with which particles react. As a consequence, the Poisson process controls the rate of transformation of calcium into gypsum in the random ordinary equations by means of the empirical particle concentration.
 
 The novelty of the model is the coupling of a system of non conservative stochastic particles  interacting both pairwise and with all the other particles by means of continuum random media.  We perform numerical experiments in order to highlight the role of the different model parameters in the evolution of the degradation. Different domains and boundary conditions are exploited.

The paper is organized as follows.  Section \ref{Sec:model} is devoted to the detailed  description of the   new hybrid model. In Section \ref{Section:qualitative}, we discuss the interplay between the different model components  and their specific roles.   In Section \ref{Sec:nondimensionalization}, we non-dimensionalize the model equations.  Finally,  Section \ref{Sec:numericalresults} is devoted to the numerical experiments. The dimensionless model system is simulated via a stochastic forward Euler-Maruyama scheme by using Julia programming language.  It turns out that the proposed hybrid  model is able to capture the main features of the corrosion process for a large variety of parameters and simulation scenarios.  We conclude the paper with Section \ref{Sec:conclusion}, where we provide a brief overview on our findings and discuss further the potential applicability of our model.

\section{The Mathematical Model}\label{Sec:model}

Let $(\Omega_P, \mathcal{F},\{\mathcal{F}_t\}_{t\in [0,T]},\mathbb{P})$ be a filtrated probability space and choose $T \in \mathbb R_+$ to be a sufficiently large time horizon. All  stochastic processes along the paper are assumed to be defined and adapted to $ \{\mathcal{F}_t\}_{t\in [0,T]}.$  Let $D\subset \mathbb R^d$, $d\geq 2$ be a bounded spatial domain. We  denote by $  \{\Delta\} \not\subset D$ the \emph{cemetery state} and let us set $\widetilde{D}=D\times \{\Delta\}.$

\paragraph{Overall dynamics}
    Let   $N\in \mathbb N$  be the number of  sulfuric acid particles located in the domain $D$ at time $t=0$. Their positions  are  represented by $\widetilde{D}$-valued stochastic processes $X^i=(X_t^i)_{t\in [0,T]}, i=1,\ldots, N$. Random particles  are Brownian diffusions subject to pairwise interactions  and to non local interactions with the surrounding environment. In the concrete application that we model, the  behavior of the particles is strongly influenced by the geometry of the environment, particularly by calcium carbonate and gypsum local densities, denoted by $c$ and $g$, respectively.
    The sulfuric acid particles undergo a chemical reaction with calcium carbonate leading to the conversion of carbonate into gypsum, during which acid particles may be  consumed and removed.  
    To model this combined effect, the motion of the sulfuric acid particles is governed by a system of stochastic differential equations (SDEs); each SDE solution evolves   until a random time in which the corresponding particle reacts and is destroyed, with the reaction events occurring according to point processes. The evolution of the underlying environmental densities   $c(t,x)$ and $g(t,x)$ follows a couple of  random ordinary differential equations (rODEs), for $(t,x)\in [0,T]\times D$, where randomness arises from the coupling with the random  empirical concentrations  of the acid particles.

       \subsection{The variables: particles and environment}
       
\paragraph{The particle processes}  The state of the $i$-th particle,  out of  $N$ initial particles of sulfuric acid, is described by a bivariate stochastic process    \begin{equation}\label{eq:particle_processes}
  (X^i,H^i)=\left(X^i_t,H^i_t \right)_{t\in[0,T]},
\end{equation}
 with state space given by  $\widetilde{D} \times \mathbb H = D\cup \{\Delta\} \times \mathbb H,$ where $\mathbb H=\{0,1\}$. \\ More precisely,   the marginal process $X^i$ models the  position in $\widetilde{D}$ of the center of the $i$-th particle. For any time $t\in   [0,T_i) $, where 
  $T_i$ is a $\mathcal{F}_t$-stopping time describing the random time when the $i$-th particle reacts with the underlying field $c$, $X_t^i$ is $D$-valued; after the reaction, i.e. for any $t\in [T_i,T]$, $X_t^i \in \{ \Delta \} \not\subset D$  and it is removed from the dynamics. As usually done, we refer to $\{\Delta\}$ as the cemetery state for the reacted particles.
  The second component is a dichotomous process which is defined, for any $t\in [0,T]$, as 
 $$
  H_t^i   := 1_{[T_i,+\infty)}(t)  .
 $$
  This process labels all the particles:  $H_t^i =0$ indicates that the $i$-th particle is still active and able to react, while $H_t^i = 1$ means that the $i$-th particle has already reacted and it is thus removed from the dynamics.

 \paragraph{Empirical measure and density}
     Let $\nu_t^N$ denote the empirical   measure associated to the active particles, i.e.
    \begin{equation}\label{eq:empirical_measure}
            \nu_t^N (dx) = \frac{1}{N}\sum_{i=1}^N   \varepsilon_{\left(X_t^i ,H^i_t\right)}\left(dx \times\{0\}\right),
        \end{equation}
    where $\varepsilon_x$ is the Dirac  measure located in $x$. Note that  $\nu_t^N(dx)$ is a spatial distribution of the only particles  still active at time $t$, therefore up to their reaction time. Then \begin{equation}\label{eq:empirical_measure_limited}
       \nu_t^N (D)\le 1,
    \end{equation}which means that  the particle system is not conservative.
    
    We consider a mollified version $u_N$  of the empirical measure, namely for any $(t,x) \in [0,T]\times D$ given by
    \begin{eqnarray}\label{def:empirical_concentration_u_N}
             u_N (t,x) &:=& u_{\text{ref}} \left( K*\nu_t^N \right) (x),
        \end{eqnarray}
    which is obtained through a convolution with a smooth estimating kernel $K\in L^\infty(D)$ and where 
    the coefficient $u_{\text{ref}}\in \mathbb R_+$ represents a characteristic mass concentration ($M/L^d$) related to sulfuric acid. We refer to $u_N$ as the \emph{empirical active particle concentration}. From \eqref{eq:empirical_measure_limited}
         and \eqref{def:empirical_concentration_u_N} we easily deduce that  $$
        u_N\in C_b\left([0,T];L^\infty(D)\right).
    $$
    The introduction of a suitable mollifier allows us to simplify the complexity of the mathematical analysis  \cite{2016_MZCJ,2012_CMF}.

\subsection{ Calcium carbonate and gypsum densities and their evolution}
    The calcium carbonate density $c$ and  the gypsum density $g$ are two  random fields on $[0,T]\times D$ evolving in time according to their mass action laws, which are described via a coupled system of rODEs, given, for any $(t,x)\in (0,T]\times D,$ by
\begin{eqnarray}\label{eq:c,g}
        \begin{cases}
        \frac{\partial}{\partial t} c (t,x) &= - \lambda\ c(t,x)\ u_N(t,x), \\[1mm]
        \frac{\partial}{\partial t} g (t,x) &= + \lambda\  c(t,x)\ u_N(t,x). 
        \end{cases}    \end{eqnarray} The parameter  $\lambda$  in \eqref{eq:c,g} is   the degradation rate for $c$.     
Therefore, although the evolution law is deterministic, randomness  arises from the   fact that the variation of calcium carbonate density  is  directly proportional to the empirical active particle concentration $u_N$,  introduced  in \eqref{def:empirical_concentration_u_N}. As a result, the environmental field densities, which evolve at the macroscale, are influenced by the microscale dynamics of sulfuric acid particles via an averaged measure of their spatial distribution.   The two random fields satisfy the  following initial conditions, for any $(t,x)\in (0,T]\times D,$ 
    $$  c(0,x) := c_0(x), \quad   g(0,x) := g_0(x),
    $$
    where $c_0,g_0: D \to \mathbb{R}_+$ are measurable, non-negative, bounded and square integrable functions, i.e.   $c_0,g_0\in \mathcal{B}^+(D) \cap L^2(D)$.
          As time elapses, calcium carbonate is gradually consumed and gypsum is produced.
\paragraph{Field properties} From \eqref{eq:c,g} we first note that, according to the law of mass action, for any time $t\in [0,T]$, we have  $c(t,x)+g(t,x)=c_0(x)+g_0(x),  x\in D.$  Furthermore, calcium carbonate can be obtained as an explicit solution as, for $(t,x)\in [0,T]\times D,$
$$c(t,x)=c_0(x)\exp\left( -\lambda \int_0^t u_N(s,x)ds \right).
$$Hence,  for any $t\in [0,T]$, the following bounds for the environmental random fields hold,
\begin{eqnarray*}
    c(t,x) &\in & \left[0,c_0(x)\right];  \\
    g(t,x) &\in & \left[g_0(x),c_0(x)+g_0(x)\right].
\end{eqnarray*}
Consequently, we may conclude that the random fields  $c,g \in {C}_b([0,T],L^2(D))$ in a pathwise sense. Finally, the $\mathcal{F}_t$-progressive measurability  of the bivarate process $(c,g)$ follow from the same measurability property of $u_N.$

\subsection{Sulfur dioxide particle dynamics } 
Sulfuric particles, described by the bivariate process \eqref{eq:particle_processes}, evolve randomly  according to a system of It\^o SDEs    till random times,  occurring in dependence of the calcium carbonate densities. More precisely  the dynamics of sulfur dioxide particles  is given by the following system, for $i=1,\ldots,N$
$$
dX^i_t=F^i(t,X_t^i,\nu^N_t,c,g) dt +\sigma dW^i_t, 
   \quad t\in [0,T_i],$$ 
   where $\{W^i\}_i$  is   a family of independent Brownian motions. In the following we describe the law of the random times $T^i$ associated with the i-th particle as well as of the drift $F^i$, modeling the interaction between particles and the effect of the environment upon particle velocity.

\paragraph{Particle chemical reaction - Particle-field interaction }
As already remark, for any $i=1,\ldots,N$,  each process $X^i$ in  \eqref{eq:particle_processes} is characterized by its lifetime $T_i$, the random time at which the $i$-th particle  randomly reacts with the environmental field $c$. 
We plan to model the occurrence of the particle chemical reaction by point  processes \cite{MACH2023_MRU,2016_MZCJ,LimLuNolen}.
Let  $\Pi$  be  a Poisson random measure  on the Borel sets $ \mathcal{B}_{[0,T]}$ of  $[0,T]$  defined as
\begin{equation}\label{def:point_process}
    \Pi(dt) := \sum_{i=1}^N \varepsilon_{T_i}(dt),
\end{equation}
so that for any $[a,b]  \in\mathcal{B}_{[0,T]} $, the random measure $ \Pi([a,b])$  counts the number of reactions occurring during $[a,b]$. Consequently, the number $N_t$ of active  particles  at time $t$ may be expressed as 
$$     N_t = N - \Pi([0,t]).$$
The   stochastic intensity measure  of the point process  is given by the following measure
\begin{eqnarray*}
    \Lambda(dt) &:=& \mathbb{P}\left(\Pi(dt) = 1|\mathcal{F}_{t^-}\right)=\lambda_t dt,
\end{eqnarray*}
 with inhomogeneous  density of the intensity measure proportional to the density of the calcium carbonate at the location of the active sulfur dioxide prticles
\begin{equation}\label{eq:point_process_rate}
    \lambda_t  := \sum_{i=1}^{N} \lambda_t  ^i= \lambda \sum_{i=1}^{N} \  c(t,x) \varepsilon_{\left(X_t^i,H_t^i\right)}(dx\times \{0\}).
\end{equation}
We may express the counting process as a sum of $N$ Poisson measures and read the times $T_i$ as the first time occurrence of the $i$-th process. More precisely, we consider  a family of standard Poisson processes
$\{\widetilde{N}^i\}_{i=1,...,N}$ which is independent both of the $N$ Brownian motions and of the particle initial conditions. The reaction time  $T_i$  of the $i$-th particle is the first (and only) jump time of the inhomogeneous Poisson process $N_t^i := \widetilde{N}^i_{\Lambda^i([0,t])}$, where 
$$ \Lambda^i\left([0,t]\right)=\int_0^t \lambda^i_s ds.
$$ Equivalently, we may say that given a random variable $Z\sim \exp(1)$, the reaction time $T_i$ is defined as
\begin{equation}\label{eq:exponential_random_time} T_i=\inf\limits_{t}\left\{ Z\le \Lambda^i\left([0,t]\right) \right\}.\end{equation}

\begin{figure}[h!]
    \centering 
    \begin{subfigure}[b]{0.49\linewidth} 
        \centering
        \includegraphics[width=\linewidth]{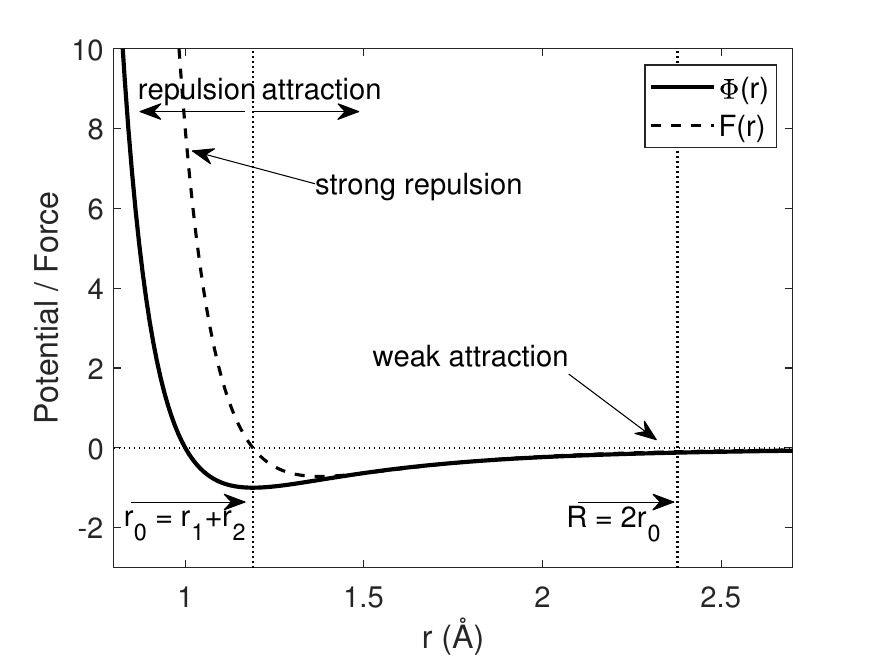} 
        \caption{}
        \label{fig:LJ} 
    \end{subfigure} 
    \hfill 
    \begin{subfigure}[b]{0.49\linewidth} 
        \centering 
        \includegraphics[width=\linewidth]{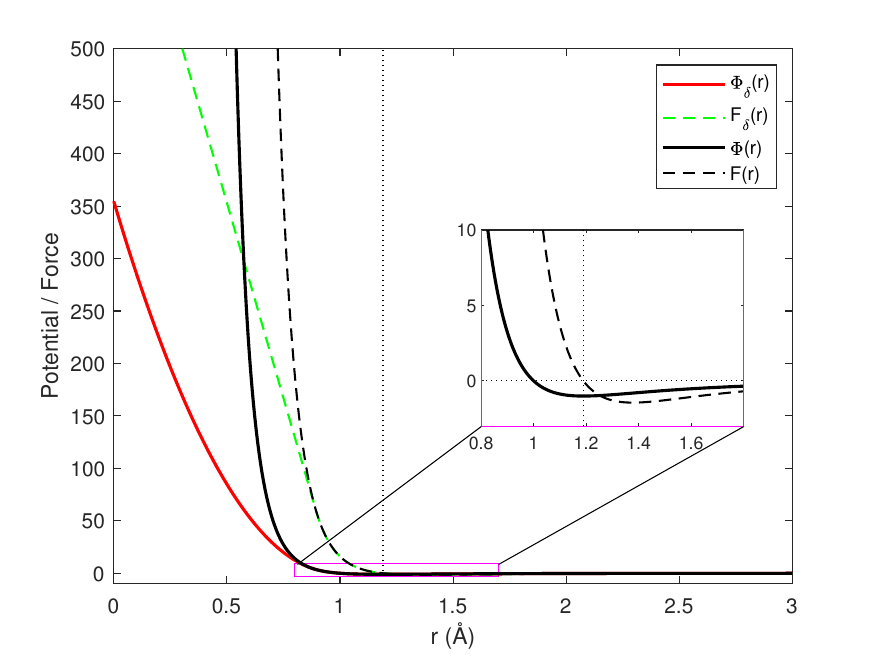}
        \caption{}
        \label{fig:LJ&smoothed} 
    \end{subfigure}
    
    \begin{subfigure}[b]{0.49\linewidth} 
        \centering 
        \includegraphics[width=\linewidth]{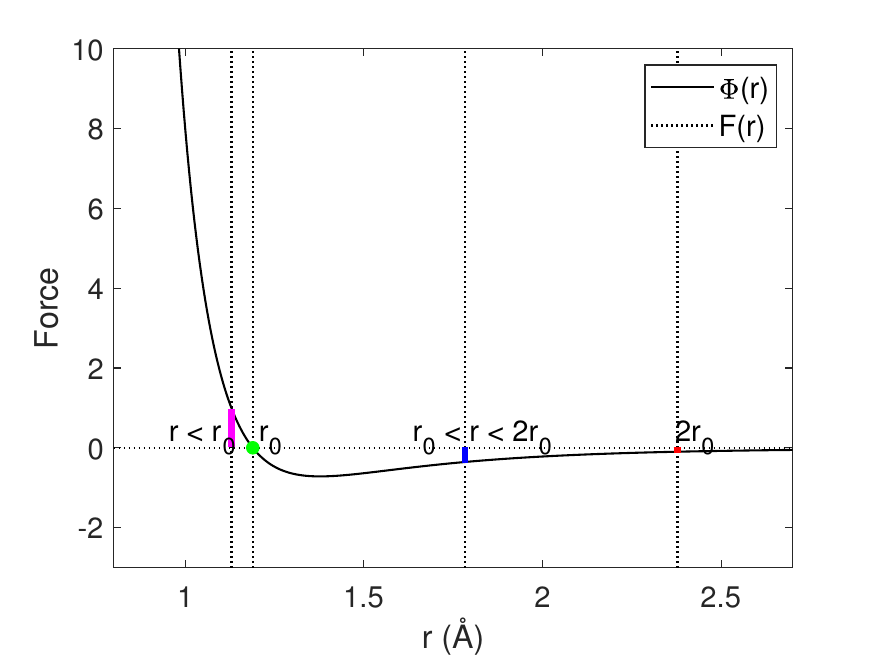} 
        \caption{}
        \label{fig:LJ_rangeplot}
    \end{subfigure}
    \hfill 
    \begin{subfigure}[b]{0.49\linewidth} 
        \centering
        \includegraphics[width=\linewidth]{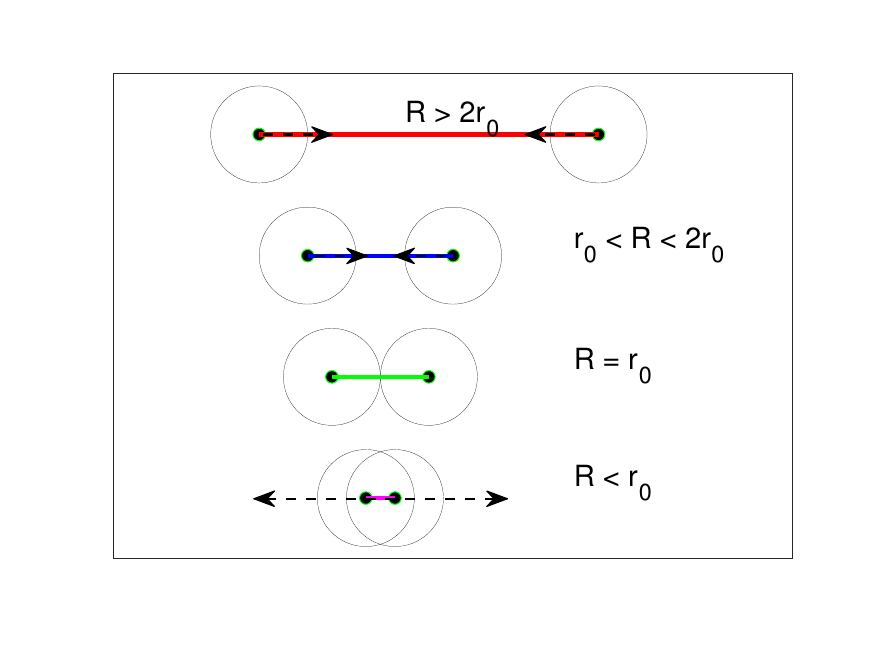}
        \caption{} 
        \label{fig:LJ_range_arrows}
    \end{subfigure} 
    
    \caption{(a) The standard Lennard-Jones potential \eqref{eq:LJ} and force in dimension \( d=3 \), with well depth $\epsilon = 1$ and range parameter $\varsigma = 1$.  
    (b) The smoothed potential \eqref{eq:LJpot_regularized} and force, with \( \delta = 0.5 \).
    (c), (d) Interaction ranges of the Lennard-Jones potential \eqref{eq:LJ}. Most of the interaction is concentrated around the equilibrium distance \( r_0 = r_1 + r_2 \), while for particles at distances greater than \( 2r_0 \) the interaction is negligible.}
   % \label{fig:LJ_combined}
\end{figure}

\paragraph{Particle-particle interaction} The interactions between particles are determined by their mutual distances, combining a strong short-range repulsion with a possible weak long-range attraction. At short distances the repulsive force dominates, preventing particles from overlapping. At larger distances, a weak attractive force is still present to account for van der Waals effects, which arise from induced dipole interactions. 
 Similarly as in \cite{2016_MZCJ,2023_FLR},  we describe the pairwise interactions via the Lennard-Jones potential \cite{LJ}. This potential captures both the repulsive and attractive forces as a function of inter-particle distance. Although typically defined in three dimensions ($d=3$), here we introduce a generalized potential valid for an arbitrary dimension $d$
\begin{eqnarray}\label{eq:LJ}
    \Phi(r) &=& 4\epsilon\left[\left(\frac{\varsigma}{r}\right)^{4d}-\left(\frac{\varsigma}{r}\right)^{2d}\right],
\end{eqnarray}
where $r$ denotes the distance between the particles, $\epsilon$ is the well depth and $\varsigma = (r_1+r_2) 2^{-1/2d}$ is the range parameter, with $r_1,r_2$ being the van der Waals radii of the interacting particles. Most of the interaction is concentrated around the equilibrium distance $r_0 = r_1+r_2$, while for particles at distances greater than $2r_0$ the interaction is negligible (see Fig. \ref{fig:LJ} and \ref{fig:LJ_rangeplot}).  
We note that the potential \eqref{eq:LJ} is strongly singular at the origin. When particles overlap the resulting force is explosive, which could create issues with well-posedness.

In order to overcome such a problem, we consider a regularized version of the Lennard-Jones potential and thus of the related force.
Specifically, we regularize the potential in an interval $[0,\delta]$ with a small $\delta>0$, by extending it with its Taylor polynomial at the second order, such that the regularization is of class $\mathcal{C}^2$. More precisely, we put
\begin{equation}\label{eq:LJpot_regularized}
    \Phi_\delta(r) := 
    \begin{cases}
        4\epsilon \left[ \left( \frac{\varsigma}{\delta} \right)^{4d} -  \left( \frac{\varsigma}{\delta} \right)^{2d} \right] - \frac{16\epsilon d}{\varsigma} \left[\left( \frac{\varsigma}{\delta} \right)^{4d+1} -\frac{1}{2}  \left( \frac{\varsigma}{\delta} \right)^{2d+1} \right] (r - \delta) \\
        \hspace{1cm} + \frac{8\epsilon d}{\varsigma^2} \left[ (4d + 1) \left( \frac{\varsigma}{\delta} \right)^{4d+2} - \frac{2d+1}{2}\left( \frac{\varsigma}{\delta} \right)^{2d+2} \right] (r - \delta)^2,
        & r\in [0,\delta]; \\[10pt]
        4\epsilon\left[\left(\frac{\varsigma}{r}\right)^{4d}-\left(\frac{\varsigma}{r}\right)^{2d}\right], & r\geq \delta.
    \end{cases}
\end{equation}
The corresponding force is therefore linear and bounded nearby the origin, which means that the singularity is removed. The related force reads as follows
\begin{equation*} 
    F_\delta(r) :=
    \begin{cases}
        \frac{16\epsilon d}{\varsigma} \left[\left( \frac{\varsigma}{\delta} \right)^{4d+1} -\frac{1}{2}  \left( \frac{\varsigma}{\delta} \right)^{2d+1} \right]\Vec{r}\\
        \hspace{1cm} - \frac{16\epsilon d}{\varsigma^2} \left[ (4d + 1) \left( \frac{\varsigma}{\delta} \right)^{4d+2} - \frac{2d+1}{2}\left( \frac{\varsigma}{\delta} \right)^{2d+2} \right] (r - \delta)\Vec{r},
        & r\in [0,\delta]; \\[10pt]
        \frac{16\epsilon d}{\varsigma} \left[\left( \frac{\varsigma}{r} \right)^{4d+1} -\frac{1}{2}  \left( \frac{\varsigma}{r} \right)^{2d+1} \right]\Vec{r}, & r\geq \delta.
    \end{cases}
\end{equation*}
Furthermore, this regularization preserves the monotonicity of the force at close ranges. In our simulations,  we clearly observe how the number of particles visiting the regularized area decreases as the parameter $\delta$ becomes small, provided that a sufficiently small time step is taken.
\noindent
The total force acting on the $i$-th particle located in $X_t^i\in D$ during the time $[t,t+dt[$ is given by  $\beta F_{part} (X_t^i)$  where
\begin{equation}\label{eq:force_particle_particle}
    F_{part} (X_t^i) := -  \sum_{ \stackrel{j:H^j_t=0}{j\neq i} }\nabla \Phi_\delta \left( \left|X_t^j -X_t^i \right| \right) \ := -\ \left(\nabla \Phi_\delta*N\nu_t^N\right) \left(X_t^i \right),
\end{equation}
 where the constant $\beta$ has unity of measure $\left[\beta \right] = TM^{-1}$ and $\Phi_\delta$ is defined in \eqref{eq:LJpot_regularized}.

\begin{figure}[h!]
    \centering
    \begin{subfigure}{0.49\linewidth}
        \centering
        \includegraphics[width=\linewidth]{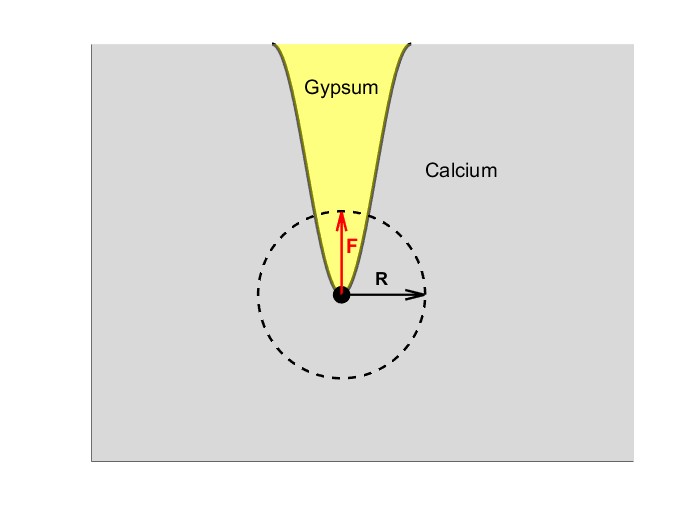}
        \caption{}
        \label{fig:f}
    \end{subfigure}
    \hfill
    \begin{subfigure}{0.49\linewidth}
        \centering
        \includegraphics[width=\linewidth]{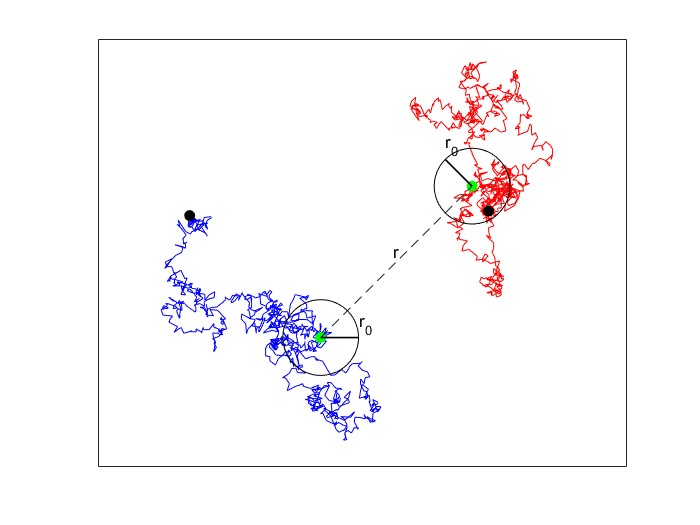}
        \caption{}
        \label{fig:BM}
    \end{subfigure}
    \caption{(a) In a heterogeneous environment, particles are attracted toward regions with higher gypsum density, \eqref{eq:force_environment}, as  the  porosity of gypsum facilitates particle diffusion. (b) Dynamics in a homogeneous environment, where motion is primarily stochastic and resembles Brownian motion.}
    %\label{fig:particles}
\end{figure}

\paragraph{Particle-field interaction} Since gypsum has significantly higher porosity than calcium carbonate, the particles are more likely to migrate toward regions with greater gypsum density, that is where more empty space is available (Fig. \ref{fig:f}). To capture this behavior, we assume that particles experience an increase in velocity toward those surroundings rich in gypsum.

A non-local interaction with the continuum fields  is introduced  via an  environment   anisotropic force  $\gamma  F_{env}(x), $  where for $x\in D$
\begin{equation}\label{eq:force_environment}
    F_{env}(x) :=  \displaystyle\int_D \frac{y-x}{\left|y-x \right|} f\left(|y-x| ,g(t,y),c(t,y)\right)\ dy.
\end{equation}
A particle located in $x$ interacts with environment locations $y\in D$ \cite{2019_FL};  the unitary vector   $(y-x)/{\left|y-x\right|}$ gives the direction of the force, while  the strength of the interaction is described by $f$ which modulates the interaction  and depends on the local environment in $y$, as follows, for any $r,v,w\in \mathbb R_+$
\begin{equation}\label{def:f}
    f\left(r;v,w\right) := \frac{v}{v + w}\ e^{-r}  1_{(0,\infty)}(v)\, 1_{(0,R]}(r),
\end{equation}
where $R>0$ denotes the sensing range of sulfuric acid for the surrounding environment. This function is effective because, in the absence of gypsum, it ensures that there is no interaction. Additionally, in a locally homogeneous environment where the gypsum fraction is uniformly distributed in all directions, the force \eqref{eq:force_environment} vanishes, reflecting a lack of directional bias. In such a  situation, distant particles behave like a Brownian motion (see Fig. \eqref{fig:BM}). The attraction becomes stronger in regions where a higher density of gypsum is present, while the exponential structure ensures that the effect is stronger at close distances.
The coefficient $\gamma$ ensures dimensional consistency with units $\left[\gamma\right] = T^{-1}L^{1-d}$.

\paragraph{The final time $T$ of the overall model } The value for the observational  time $T$ should be sufficiently large  in order to capture the most of the overwhole dynamics. In particular, slower is the reaction, that is for smaller  $\lambda$, larger need to be $T$.
 
\subsection{The complete system   (P)}
In conclusion, we propose that the overall dynamics is described by the following SDE-rODE system
\begin{equation}\label{eq:complete_ODE_SDE}
    \begin{cases}
       \, dX_t^i &= \beta F_{part}(X_t^i)\ dt +  \gamma F_{env}(X_t^i)\ dt +  \sigma dW_t^i, \ \ \ t\in (0,T_i),  \\
        \displaystyle\frac{\partial}{\partial t} c(t,x) &= -\lambda\ c(t,x)\  u_N(t,x), \quad (t,x) \in (0,T)\times D,\\[2mm]
        \displaystyle\frac{\partial}{\partial t} g(t,x) &= +\lambda\ c(t,x)\ u_N(t,x), \quad (t,x) \in (0,T)\times D,
    \end{cases}
\end{equation}
where $i\in\{1,\dots,N\}$, the drift functions $F_{part}$ and $F_{env}$ are the interaction force acting on the $i$-th particle due the other particles and with the environment, given by \eqref{eq:force_particle_particle} and \eqref{eq:force_environment}, respectively,
and $u_N$ is the regularized empirical measure associated to the active particle system introduced in \eqref{def:empirical_concentration_u_N} starting from the empirical measure defined in \eqref{eq:empirical_measure}. 
\noindent
The system \eqref{eq:complete_ODE_SDE} is coupled with the chemical reaction, described by the point process \eqref{def:point_process} having a stochastic intensity with rate \eqref{eq:point_process_rate}. Hence, particles are active and they move till a random time defined by \eqref{eq:exponential_random_time}.
\noindent
The initial conditions for the complete system are
\begin{equation}\label{eq:IC}
    \begin{cases}
        X^i_0 &= X_I^i\qquad i.i.d., \text{not overlapping};\\
        c(0,x) &= c_0(x) \qquad x\in  {D}, \\
        g(0,x) &= g_0(x) \qquad x\in {D},
    \end{cases}.
\end{equation} 
We refer to the complete proposed model described by \eqref{eq:complete_ODE_SDE}-\eqref{eq:IC} endowed with the subsequent specifications as problem $(P)$.
\begin{remark}
 For describing the acid particles dynamics via a first order SDE a  high friction hypothesis has been assumed. This approximation is known as Smoluchowski approximation (\cite{Gro_kinsky_2004}).   
\end{remark}

Table \ref{Tab:Dimensions} shows the dimension of all the variable and parameters involved in the model.

\begin{table}[ht]
    \centering
    \begin{tabular}{|p{1.8cm}|p{6.5cm}|p{2cm}|}
    \hline
    Notation & Name of the Object & Dimension \\
    \hline
    $x$ & Position & $L^d$ \\
    $t$ & Time & $T$ \\
    $\sigma$ & Diffusion coefficient & $L^d T^{-1/2}$ \\
    $\beta$ & Particle-particle interaction coefficient & $L^2T^{-1}$ \\
    $\gamma$ & Particle-field interaction coefficient & $T^{-1}L^{1-d}$ \\
    $c$ & Calcium carbonate density & $ML^{-d}$ \\
    $g$ & Gypsum density & $ML^{-d}$ \\
    $u$ & Regularized counting measure & $ML^{-d}$ \\
    $\lambda$ & Reaction rate & $L^{d}T^{-1}M^{-1}$ \\
    $N$ & Number of particles & $\left[-\right]$ \\
    \hline
    \end{tabular}
    \caption{Physical dimensions of the variables involved in the system dynamics.}
    \label{Tab:Dimensions}
\end{table}

\section{How do the components of \textit{(P)} influence the overall dynamics?}\label{Section:qualitative}
In this section  we discuss into details the influence on the sulfuric acid particles of each term of the  complete system $(P)$ and how we expect the model behaves.

\smallskip

The Lennard-Jones interaction \eqref{eq:LJ} affects particles at close range, around the equilibrium distance, given by the sum of the particles van der Waals radii. Beyond twice this distance, particle-particle interactions are negligible (Fig. \ref{fig:LJ_rangeplot}, \ref{fig:LJ_range_arrows}).  Consequently, when particles are sufficiently far apart and the surrounding environment is homogeneous, their motion is primarily governed by the Brownian motion, leading to a diffusive behavior  \eqref{fig:BM}, where particles undergo random, isotropic and uncorrelated displacements described by the Wiener process. As particles approach each other, the dynamics becomes more complex due to Lennard-Jones inter-particle interactions \eqref{fig:LJ_rangeplot}. This potential is sharply repulsive at very short distances, where the particles are partially overlapping and weakly attractive at intermediate distances due to van der Waals forces.
This repulsion effectively makes particles act as barriers to the motion of each other, particularly in dense regions, leading to correlated motion and collective behavior.

The second term in the drift of SDE in system \eqref{eq:complete_ODE_SDE} models particle interaction with the environment \eqref{eq:force_environment}, by introducing a spatial bias governed by the function $f$ given  by\eqref{def:f}. This term reflects the tendency of particles to migrate toward areas with a higher gypsum density, where higher porosity facilitates particle diffusion compared to calcium carbonate. So the effect of this term induces a local attraction toward gypsum-rich regions (Fig. \ref{fig:f}). However, the stochastic term continues to drive some particles into contact with calcium carbonate, where they trigger chemical reactions that gradually reshape the material landscape.
Gypsum is produced in the location where the particle density is positive and the growth is proportional to the calcium carbonate density, the gypsum-covered surface area expands, particularly along the borders of pre-existing gypsum patches where particles are attracted. This creates a feedback loop: particles preferentially move along pathways where gypsum has already formed, further concentrating particle activity at these expanding boundaries. 
Finally, sulfate  particles  are active and moves till a random time defined by \eqref{eq:exponential_random_time}, when their reaction occur.

In conclusion, the interplay between random motion, gypsum attraction, and chemical reactions lead to a self-organizing system. We do expect that  gypsum pathways become more prominent, guiding the movement of the particles toward less restrictive regions and that over time, the system should exhibit heterogeneous surface coverage, with gypsum-rich areas expanding while calcium carbonate regions gradually diminish. This highlights how environmental interactions shape both particle dynamics and material transformation.

\section{Non-dimensionalization}\label{Sec:nondimensionalization}
We briefly recall the effects of a general space-time random transformation of a stochastic differential equation (see, e.g.,  \cite{2016_DVMU} for more details). We say that a couple $(X,W)$ solves the SDE $(\mu,\sigma)$, with $\mu: \mathbb{R}^d \to \mathbb{R}^d, \sigma:\mathbb{R}^d\to \mathbb{R}^{d\times m}$ being smooth functions, if $(X_t)_{t\in\mathbb{R}_+}$ is a stochastic process with values in an open subset $A\subset \mathbb{R}^d$ and W is an $m$-dimensional Brownian motion such that
\begin{equation*}
    dX_t = \mu(X_t)dt + \sigma(X_t)dW_t.
\end{equation*}
A space transformation can be represented by a diffeomorphism $\Psi: \mathbb{R}^d \to \mathbb{R}^d$ and the more general random time transformation  by a continuous strictly increasing process $t^\prime := \beta_t := \int_0^t \eta(X_s)ds$, where $\eta : M \to \mathbb{R}_+$ is a strictly positive smooth function.
The transformed process under the action of $T=(\tilde \Psi,\eta)$ can be defined as $P_T(X,W) := \left(\Psi(H_\eta(X)), H_\eta(W^\prime)\right)$, where $H_\eta$ is the operator that evaluates a process in the new time scale, i.e. $ H_\eta (W)=W_{\alpha_t}$, with $\alpha_t$ the inverse of $\beta_t.$ For time transforming the Brownian motion we need to consider  the following invariance property. Given the new process $W^\prime_{t}$ solution to the equation
\begin{equation*}
    dW^\prime_{t} = \sqrt{\eta(X_{t})} dW_{t},
\end{equation*}
one has that $H_\eta (W^\prime)$ remains a Brownian motion \cite[Th. 7.20]{SDE}.
The whole transformed process is solution of the transformed SDE $(\mu^\prime,\sigma^\prime)$ with
\begin{equation*}
    \mu^\prime := \left(\frac{1}{\eta} L(\Psi)\right) \circ \Psi^{-1},\qquad 
    \sigma^\prime := \left(\frac{1}{\sqrt{\eta}} D(\Psi) \sigma\right) \circ \Psi^{-1},
\end{equation*}
where $L$ is the differential operator $L = \left(\frac{1}{2}\sigma^T\sigma\right)^{ij}\partial_i\partial_i + \mu^i\partial_j$ and $D(\Psi)$ is the matrix with entries $ D\Psi^l_i=\partial_i\Psi^l.$
\noindent

In order to non-dimensionalize problem ($P$), we choose a set of reference quantities and  rescale time and space as 
\begin{equation*}
    t^\prime :=\frac{t}{t_{\text{ref}}}, \quad x^\prime  := \frac{x}{x_{\text{ref}}},
\end{equation*}
with $(t^\prime,x^\prime) \in [0,T^\prime]\times D^\prime$, where $T^\prime =T/t_{\text{ref}}$ and $D^\prime = \{ x^\prime = x/x_{\text{ref}} : x\in D\}$. The scaled reaction rate and fields are 
\begin{equation*}
   \lambda^\prime:=\frac{   \lambda}{   \lambda_{\text{ref}}}, \quad  
    c^\prime:=\frac{c}{c_{\text{ref}}},\quad 
      g^\prime  := \frac{g}{g_{\text{ref}}}, \quad 
    u^\prime    := \frac{u}{u_{\text{ref}}}.\\
\end{equation*}

\paragraph{Scaling the SDEs}
According with the previous notations, since $ \Psi(X) = {X}/{x_{\text{ref}}}$  from   \eqref{eq:complete_ODE_SDE} 
the rescaled SDE becomes
\begin{eqnarray*}
    dX^\prime_{t^\prime} = \frac{t_{\text{ref}}}{x_{\text{ref}} }\mu(X^\prime_{t^\prime})dt^\prime + \frac{\sigma\ \sqrt{t_{\text{ref}}}}{x_{\text{ref}} } dW^\prime_{t^\prime},
\end{eqnarray*} and we denote by $\nu_t^{\prime,N}$  the scaled empirical measure given by
\begin{eqnarray*}
    \nu_t^{\prime,N}(dx^\prime) := \frac{1}{N}\sum_{i=1}^N \varepsilon_{\left(X^{\prime,i}_{t^\prime},H^{i}_{t^\prime}\right)}(dx^\prime\times\{0\}).
\end{eqnarray*}
By expanding the drift terms, the particle-interaction force becomes
\begin{eqnarray*}
    F^\prime_{part}(X^{\prime,i}_{t^\prime}) = - \frac{t_{\text{ref}}}{x_{\text{ref}}^2} \left(\nabla_{x^\prime} \Phi^\prime_\delta * \nu_t^{\prime,N}\right) (X^{\prime,i}_{t^\prime}),
\end{eqnarray*}
while  the field-interaction term becomes
\begin{eqnarray*}
    F^\prime_{env}(X^{\prime,i}_{t^\prime}) = t_{\text{ref}} \ x_{\text{ref}}^{d-1} \displaystyle\int_{D^\prime} \frac{y^\prime-X^{\prime,i}_{t^\prime}}{|y^\prime-X^{\prime,i}_{t^\prime}|} \ f^\prime\left(|y^\prime-X^{\prime,i}_{t^\prime}| ,g^\prime(t^\prime,y^\prime),c^\prime(t^\prime,y^\prime)\right) dy^\prime\ dt^\prime.
\end{eqnarray*}
Here $\Phi^\prime_\delta$ and $f^\prime$ are simply the evaluations of $\Phi$ and $f$ in the new variables.

\paragraph{Scaling the rODEs}
The equations governing the evolution of calcium carbonate and gypsum become
\begin{eqnarray*}
   \frac{\partial}{\partial t^\prime}c^\prime (t^\prime,x^\prime) &=& - t_{\text{ref}}\  u_{\text{ref}}\ \lambda_{\text{ref}}\ \lambda^\prime c^\prime(t^\prime,x^\prime)\  u^\prime_N(t^\prime,x^\prime),  \nonumber\\\frac{\partial}{\partial t^\prime}g^\prime(t^\prime,x^\prime) &=& \frac{c_{\text{ref}}}{g_{\text{ref}}}\ t_{\text{ref}}\ u_{\text{ref}}\ \lambda_{\text{ref}}\ \lambda^\prime\ c^\prime(t^\prime,x^\prime)\ u^\prime_N(t^\prime,x^\prime).\nonumber
\end{eqnarray*}

\paragraph{Scaling the chemical reaction term}
Scaling the point process results in
\begin{equation*}
    \begin{cases}
        &\Pi^\prime (dt^\prime) = \sum_{i=1}^N \varepsilon_{T^\prime_i}(dt^\prime);\\
        &\Lambda^\prime (dt^\prime)= c_{\text{ref}}\ t_{\text{ref}}\ \lambda_{\text{ref}} \lambda^\prime N\int_D c^\prime(t^\prime,x^\prime) \nu^{\prime,N}_t(dx^\prime) \,dt^\prime.
    \end{cases}
\end{equation*}

\paragraph{The scaled system}
To normalize the particle-particle interaction and the rate of the Poisson process we choose the space-time scale
\begin{eqnarray*}
    t_{\text{ref}} &:=& \frac{1}{\lambda_{\text{ref}} c_{\text{ref}}};\qquad 
    x_{\text{ref}} := \frac{1}{\sqrt{\beta\lambda_{\text{ref}} c_{\text{ref}}}},
\end{eqnarray*}
and then define the dimensionless parameters
\begin{eqnarray*} 
    \overline{\gamma} &:=& \frac{\gamma}{\beta^{d/2}(\lambda_{\text{ref}} c_{\text{ref}})^{(d+1)/2}}, \quad \overline{\sigma} := \sigma \sqrt{\beta} \quad
    \lambda_c := \frac{u_{\text{ref}}}{c_{\text{ref}}}, \quad \lambda_g := \frac{u_{\text{ref}}}{g_{\text{ref}}}.
\end{eqnarray*}

To simplify the notation, we omit the primes in the variables. The non-dimensionalsed system $(P^\prime)$  takes then the form:
\begin{eqnarray}\label{eq:nondimensionalized_system}
    \begin{cases}
         dX^i_{t} &= - \left(\nabla_{x} \Phi * N\nu_t^{N}\right) (X^i_{t})dt +  \overline{\gamma} \displaystyle\int_{D} \frac{y-X^i_t}{|y-X^i_{t}|}  
        f\left(|y-X^i_t| ,g,c\right) dy\ dt   \\    &\,\,\,\, + \overline{\sigma} dW_{t}^i,   \hspace{4.5cm} t\in (0,T_i),\ i \in \{1,\dots, N\}\\[1mm]
         \displaystyle\frac{\partial}{\partial {t}} c(t,x) & = - \lambda_c \lambda\ c(t,x)\ u_N(t,x), \quad (t,x)\in [0,T]\times D,\\[2mm]
        \displaystyle \frac{\partial}{\partial {t}} g(t,x)& = \lambda_g \lambda\ c(t,x)\ u_N(t,x), \quad (t,x)\in [0,T]\times D,\\
        \Lambda(dt)& =\lambda\ N\int_D c(t,x) \ \nu^N_t(dx) dt \quad  \in \mathcal{M}([0,T]).
    \end{cases}
\end{eqnarray}
The system is endowed with the corresponding initial and boundary conditions.

Table \ref{Tab:Reference} shows the main reference quantities considered in the following.

\begin{table}[ht]
    \centering
    \begin{tabular}{|p{1.6cm}|p{4cm}|p{1.6cm}|}
    \hline
    Notation & Reference quantities & Units \\
    \hline
    $x_{\text{ref}}$ & $\text{diam}(D)$ & cm \\
    $t_{\text{ref}}$ &  $1/(\lambda u_{\text{ref}})$ & days \\
    $c_{\text{ref}}$ &  $c_0^{max}$ & g/cm$^3$ \\
    $g_{\text{ref}}$ &  $\overline{g}(c_{\text{ref}})$ & g/cm$^3$ \\
    $u_{\text{ref}}$ & $\overline{u}(c_{\text{ref}})$ & g/cm$^3$ \\
    $\lambda_{\text{ref}}$ & 1 & cm$^3$/(g s) \\
    \hline
    \end{tabular}
    \caption{Reference quantities for the non-dimensional system. The calcium carbonate density is normalized with its maximum initial value. The quantity $\overline{g}$ represents the sulfuric acid mass concentration required to fully dissolve the initial calcium carbonate, making it proportional to $c_{\text{ref}}$. The parameter $\overline{u}$, derived from empirical data, is related to sulfuric acid properties.}
    \label{Tab:Reference}
\end{table}

\section{Numerical experiments}\label{Sec:numericalresults}
We consider the non-dimensionalized system \eqref{eq:nondimensionalized_system} on  $ [0,T]\times D,$  where $D=[0,1]^2$ and $T=1$.   
The spatial domain is discretized using a $100\times 100$ grid with a spatial step $\Delta x = 10^{-2}$, and we employ a time step $\Delta t = 5\times10^{-6}$. The choice of the parameters in \eqref{eq:nondimensionalized_system} ensures to be at the scale of the Lennard-Jones interaction.  Moreover, for simplicity and to better understand the intrinsic behavior of the system, all reference quantities ($c_{\text{ref}},g_{\text{ref}}$ and $u_{\text{ref}}$) are set to 1. However, in applications to real-world system, these values should be estimated from physical data.
Table \ref{tab:parametervalues} shows the parameters related to time, space and dimension and interaction of particles. 

\begin{table}[h!]
    \centering
    \begin{tabular}{|c|c|c|}
             \hline
         Parameters& Definition& Value \\
         \hline
         d& space dimension& 2\\
          dx& spatial step& 0.01\\
         dt& time step& $5\times 10^{-6}$\\
        n& particles number& 100\\
         r & particles radius& 0.01\\
         $\epsilon$ & potential well depth & 1 \\
         $\delta$ & regularization parameter & $0.5 r$ \\
         R & range parameter& $4 r$ \\ 
                  \hline
    \end{tabular}
    \caption{Dimensionless space, time and particle-related parameters.}
    \label{tab:parametervalues}
\end{table}

\subsection{Exploring different initial and boundary conditions}

 The first experiment we carry out aims to explore the time evolution of the system in dependence of different initial and boundary conditions.  The kinetic parameters in Table \ref{tab:kinetic_parameters_experiment1} are used. 
\begin{table}[h!]
    \centering
    \begin{tabular}{|c|c|c|}
             \hline
         Parameters& Definition& Value \\
         \hline
         $ \overline{\gamma}$& particle-field coefficient& 100\\
         $\overline{\sigma}$ & diffusion coefficient& 0.1\\
          $\lambda$& reaction rate &$5\times10^3$\\
                  \hline
    \end{tabular}
    \caption{First experiment: kinetic parameters.}
    \label{tab:kinetic_parameters_experiment1}
\end{table}
We consider three different initial configurations to explore how spatial distributions of reactants influence the evolution of the system, i.e. how reaction pathways emerge under different geometric constraints.  
\begin{itemize}
    \item Case 1:  a block of calcium carbonate is exposed to sulfuric acid particles only from the top, simulating a directional attack. Periodic boundary conditions apply.  
    \item Case 2:    calcium carbonate is surrounded by sulfuric acid particles on all four sides.  Periodic boundary conditions apply.  
    \item Case 3:  we simulate a possible damage;   an indentation in the calcium carbonate block is considered, where acid particles accumulate and initiate localized corrosion.  Periodic boundary conditions apply.  
    \item Case 4. a block of calcium carbonate is exposed to sulfuric acid particles only from the top, simulating a directional attack.  Top and bottom reflecting boundary condition are considered.
\end{itemize} 

In all these cases, the initial conditions for calcium carbonate and gypsum are defined on disjoint subsets of $D$. The acid particles have random initial positions within the region where calcium carbonate is absent, i.e. where $c_0(x) = 0$ for $x\in D$. To ensure adequate spacing between particles at $t=0$, the initial area not covered by calcium carbonate have been kept roughly the same across the different cases. However the extent of the calcium carbonate surface exposed to acid attack has been allowed to vary. To follow the overall model dynamics over time, we plot the density maps for calcium and gypsum at specific times, where the positions of sulfuric acid particles are represented by green balls that get replaced with red crosses after they react and are consumed.

\subsubsection{Case 1: One-sided corrosion}
The initial conditions for environmental fields are, for any $x\in D.$
\begin{equation*}
    g_0(x) = \ 0;  \quad 
    c_0(x) =  1_{[0,1]\times[0,0.8]}(x).
\end{equation*}
At $t=0$, particles are uniformly distributed outside the calcium carbonate block, that is
\begin{equation*}
        X_0^i \sim U([0,1]\times[0.8,1]).
\end{equation*}
Periodic boundary conditions are considered. This scenario corresponds to a block of calcium carbonate exposed to acid particles on the top surface, where gypsum has not yet formed.

\begin{figure}[h!]
    \centering
    \includegraphics[width=0.9\linewidth]{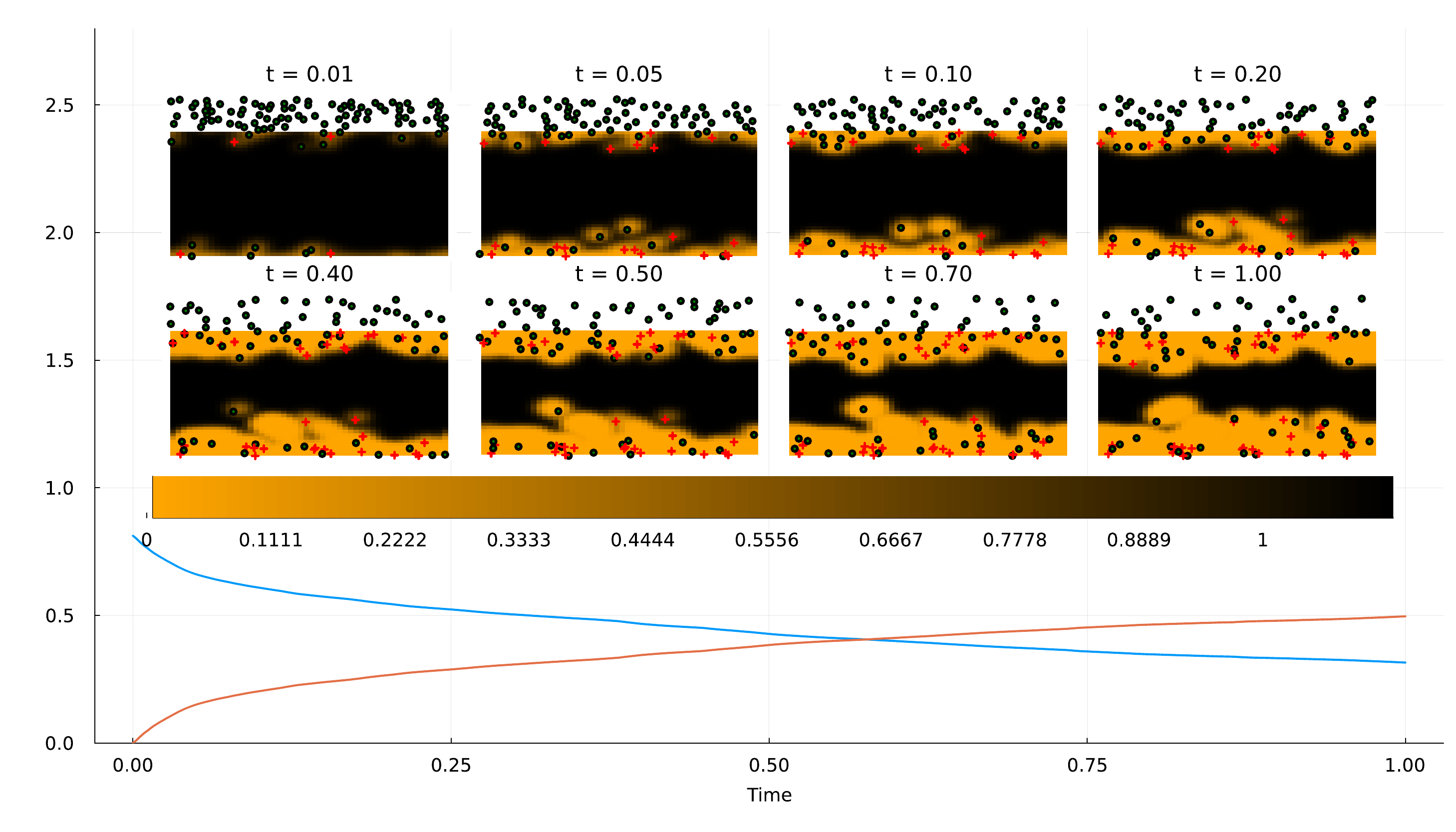}
    \caption{Case 1.  (Top) Evolution of the system at specific times: spatial  gypsum (orange) and calcium (black) densities; active  (green circle)  and reacted (red cross)   sulfuric acid particles locations.  (Bottom) Temporal evolution of the total mass mass of gypsum (solid light blue) and calcium carbonate (orange).}
    \label{fig:Case1}
\end{figure}
\begin{figure}[h!]
    \centering
    \includegraphics[width=0.9\linewidth]{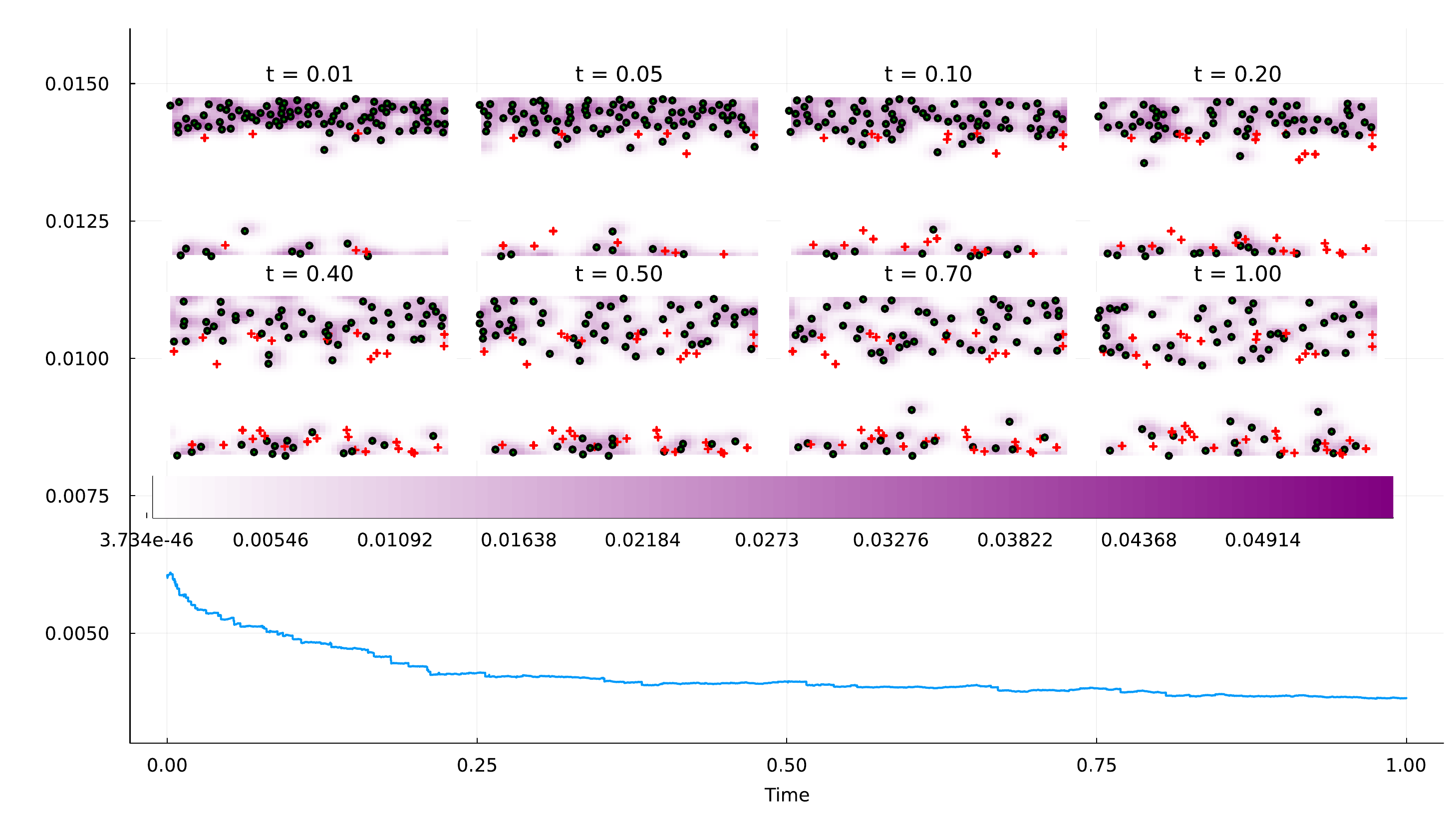}
    \caption{Case 1. (Top) Evolution of particles concentrations at specific times. (Bottom) Temporal evolution of the total mass of sulfuric acid particles.}
    \label{fig:Case1_U}
\end{figure}

Figure \ref{fig:Case1} shows the progression of the evolution. The total mass of gypsum becomes larger as time increases  as shown by the picture at the bottom, where the total mass of the fields are shown.   The observed  heterogeneity of the spatial distribution of the underling fields comes from randomness induced by the coupling with the stochastic empirical concentration of sulfuric acid particles $u_N(t,x)$.  Green spheres represent sulfuric acid particles, which move dynamically within the domain as time progresses.
When these particles encounter calcium carbonate (black regions), they react and are consumed, as indicated by red crosses marking their final positions. As introduced in Section \ref{Section:qualitative}, acid particles tend to migrate toward gypsum-rich regions due to the high porosity of the material, moving around the pathways opened by its formation, but stochastic motion continuously drives some particles into contact with calcium carbonate over time, sustaining the reaction. 

 Figure \ref{fig:Case1_U}   illustrates the evolution of sulfuric acid particles and their concentration field $u_N(t,x)$. What is worth noting here is that particle consumption is an instantaneous event, where particles loose all their mass at once when they react. Whether a particle is consumed by the reaction or not, its lifetime is governed by a Poisson process.

\subsubsection{Case 2: Four-sided corrosion}
\begin{figure}[h!]
    \centering
    \includegraphics[width=0.9\linewidth]{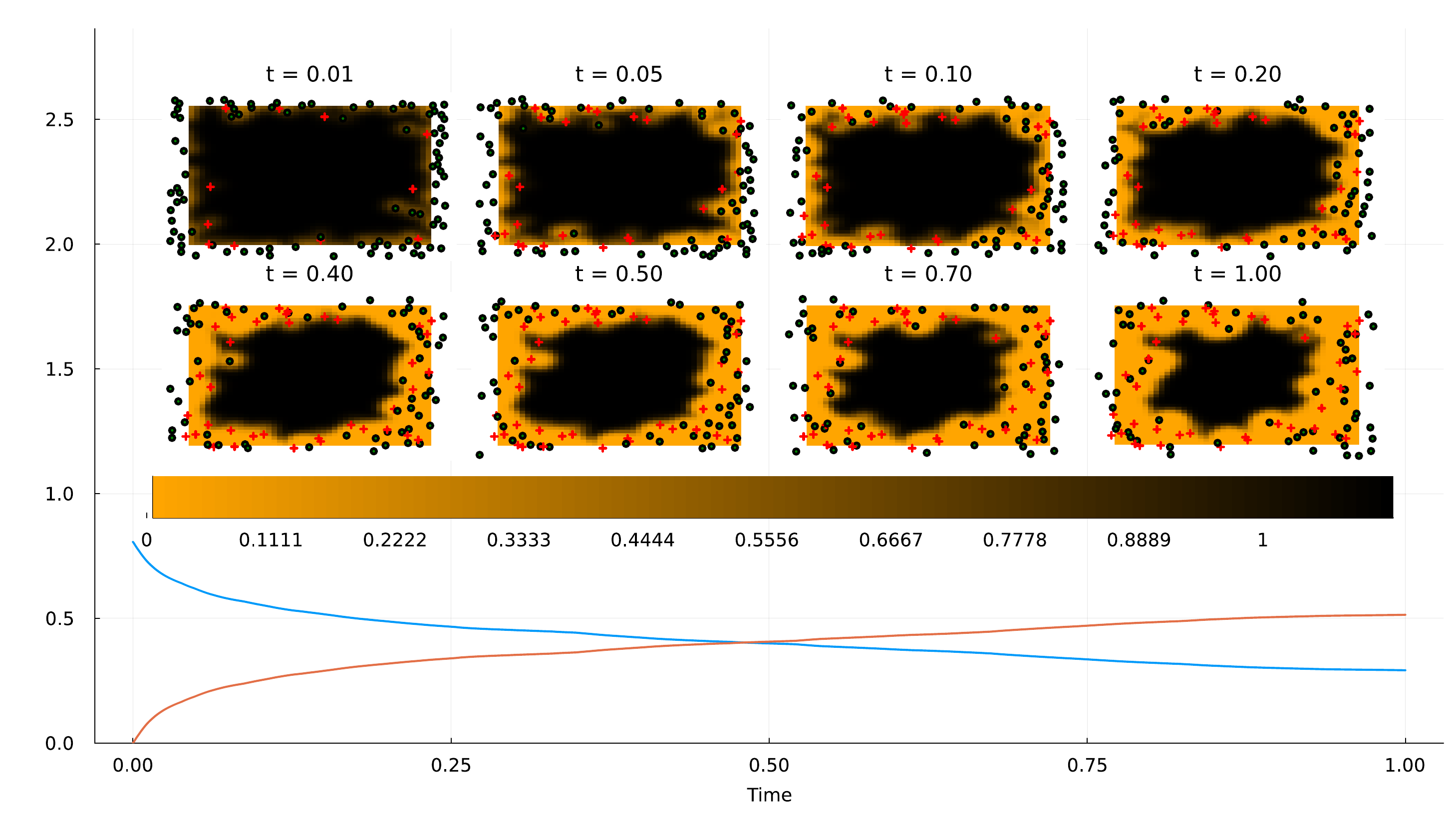}
    \caption{Case 2.  (Top) Evolution of the system at specific times: spatial  gypsum (orange) and calcium (black) densities; active  (green circle)  and reacted (red cross)   sulfuric acid particles locations.  (Bottom) Temporal evolution of the total mass mass of gypsum (solid light blue) and calcium carbonate (orange).}
    \label{fig:Case2}
\end{figure}

Here, the initial conditions for ODEs are, for any $x\in D$
\begin{align*}
    g_0(x) = &\ 0 \quad x\in D,\\
    c_0(x) = &  1_{[0.1,0.9]\times [0.1,0.9]}(x).
\end{align*}
The sulfuric acid particles are once again uniformly distributed outside the calcium carbonate block
\begin{equation*}
        X_0^i \sim U(D\backslash [0.1,0.9]^2).
\end{equation*}
Again periodic boundary conditions are considered. This case corresponds to a scenario where a block of calcium carbonate is exposed to sulfuric acid on all sides, with an initially absent gypsum. 
In Figure \ref{fig:Case2}, we  display similarly to Case 1 the mass evolution and density maps for gypsum and calcium carbonate. Interestingly, the consumption of calcium carbonate is faster in Case 2 compared to Case 1. This should be expected since the interface is larger. The preference of acid particles to move towards gypsum-rich regions is again evident from the density maps, as seen in previous case.

\subsubsection{Case 3: Corrosion along an indentation}
\begin{figure}[h!]
    \centering
    \includegraphics[width=0.9\linewidth]{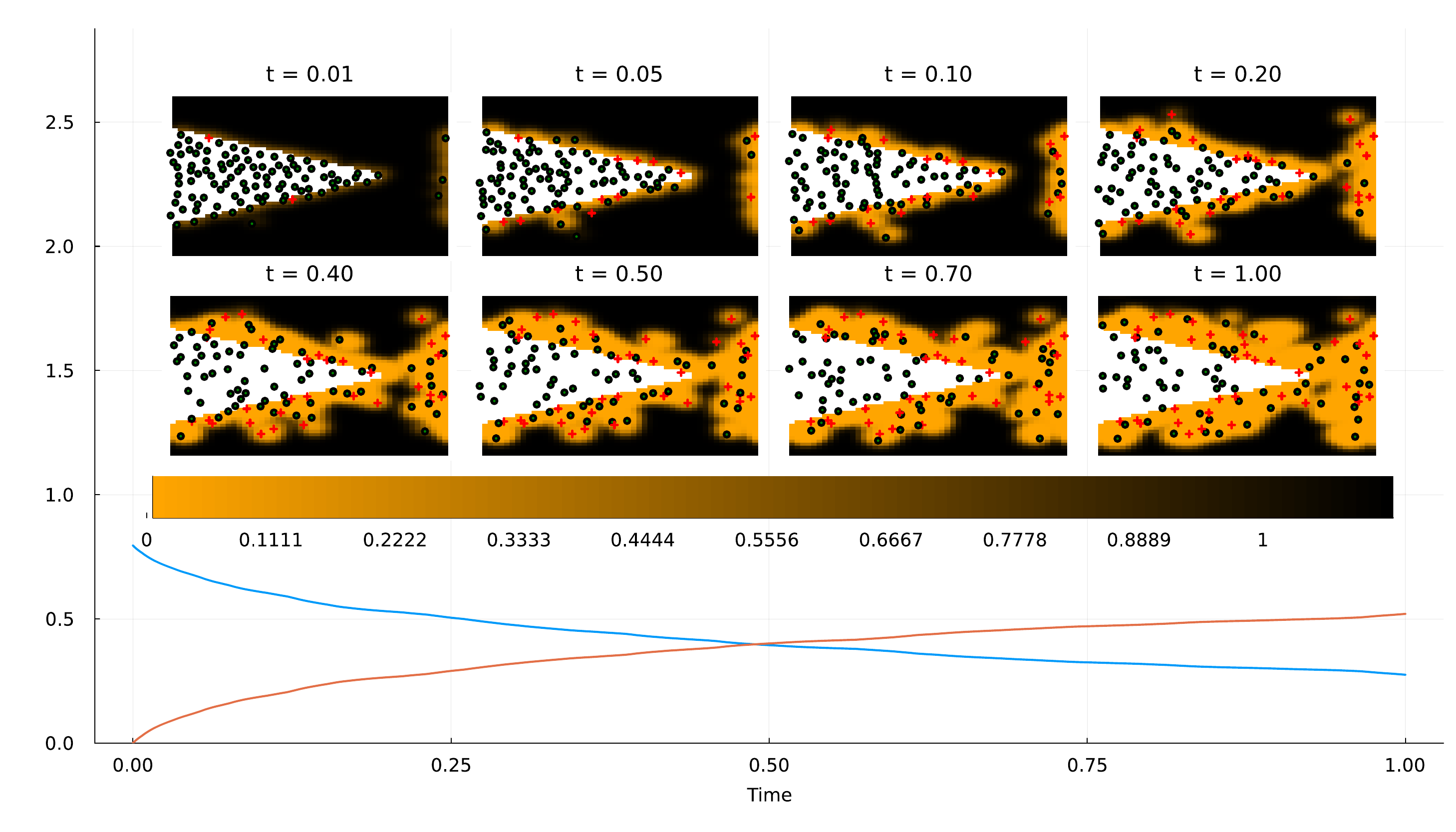}
    \caption{Case 3.  (Top) Evolution of the system at specific times: spatial  gypsum (orange) and calcium (black) densities; active  (green circle)  and reacted (red cross)   sulfuric acid particles locations.  (Bottom) Temporal evolution of the total mass mass of gypsum (solid light blue) and calcium carbonate (orange).}
    \label{fig:Case3}
\end{figure}

Now we consider a different case. We model a damage of the carbonate block given by a   triangular region $\widetilde{D}$, determined by the points $(0, 0.2),(0, 0.8),$ and $(0.8, 0.5)$. Hence, the initial conditions for this case are, for any  $ x\in D$
\begin{align*}
    g_0(x) = & 0;  \\
    c_0(x) =&  1_{\widetilde{D}}(x);
\end{align*}
and sulfuric acid particles are uniformly distributed in $\widetilde{D}$, i.e.
\begin{equation*}
        X_0^i \sim U( \widetilde{D}).
\end{equation*}
As a consequence acid corrosion is concentrated on an indentation in a block of calcium carbonate. Again periodic boundary conditions apply.
In Figure \ref{fig:Case3}, we illustrate as before the mass evolution and spatial distributions of gypsum and calcium carbonate.
The consumption rate of calcium carbonate in this case is middling when comparing with the two other cases fitting to the idea that the measure of the interface is important when all other parameters is the same. 

\begin{figure}[h!]
    \centering
    \includegraphics[width=0.9\linewidth]{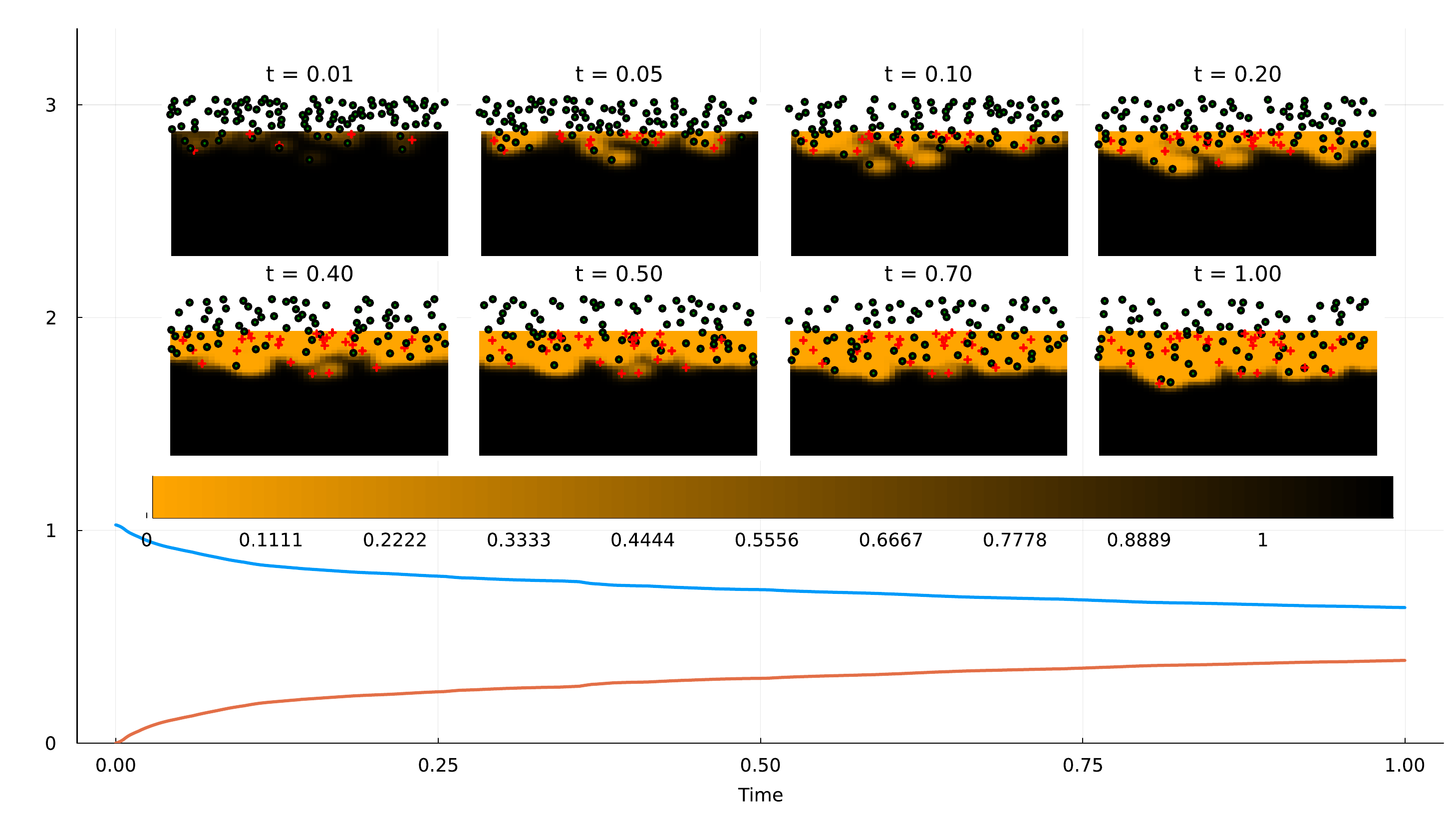}
    \caption{Case 4.  (Top) Evolution of the system at specific times: spatial  gypsum (orange) and calcium (black) densities; active  (green circle)  and reacted (red cross)   sulfuric acid particles locations.  (Bottom) Temporal evolution of the total mass mass of gypsum (solid light blue) and calcium carbonate (orange). }
    \label{fig:Case4}
\end{figure}

\subsubsection{Case 4: One-sided corrosion with reflective boundary conditions}

In this case we want to simulate the case  in which a calcium carbonate block is placed in a chamber where sulfur dioxide come only from above and not from the bottom. This means to simulate a vertical cross-section of calcium carbonate being attacked by sulfuric acid coming from above.  Hence,  instead of periodic boundary conditions, we apply reflective boundaries at the top and bottom, while keeping the   periodicity at the left and right boundaries. 
As the initial conditions regard, we consider the same ad in Case 1. 
The results are displayed in Figure \ref{fig:Case4}.
A significant reduction in the production by chemical reaction is observed; this is likely due to particle crowding. The Lennard-Jones interaction prevents particles from passing too easily each other. The reflective boundary further restricts movement. Additionally, since the top boundary is reflective, the reaction interface is halved compared to Case 1,  slowing further the calcium carbonate consumption.

\subsection{Case 1: a parameter study}
\begin{figure}[h!]
    \centering
    \includegraphics[width=0.8\linewidth]{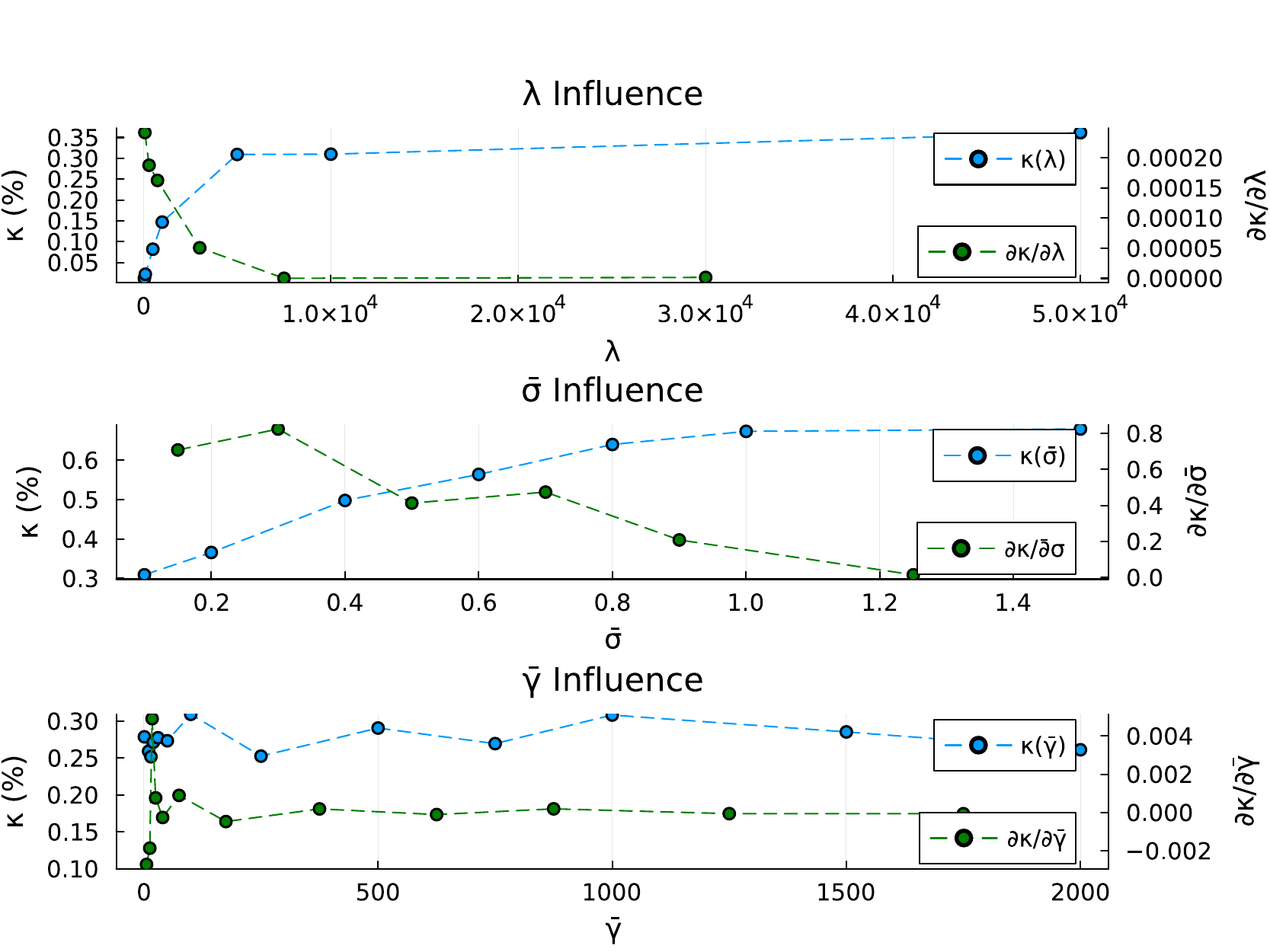}
    \caption{Percentage of consumed calcium carbonate   with its numerical derivatives of the consumed percentage with respect to each kinetic parameter.}
    \label{fig:parameterplot}
\end{figure}

In the previous experiments, we have considered the same kinetic  parameters, and only changed the initial and boundary conditions.  In this section, we focus only on Case 1 and quantify how the three interaction parameters, $ \overline{\sigma},\overline{\gamma}$, and $\lambda$ affect the corrosion of calcium carbonate.  We denote by $\kappa$ the percentage of calcium carbonate consumed 
$$
\kappa=\frac{\int_D c(T,x)dx}{\int_D c(0,x)dx}.
$$ Simulation are performed in $[0,T]=[0,0.1].$ 
Table \ref{tab:parameterexploration} shows the different configuration of the kinetic parameters, along to the corresponding value of  $\kappa$, while  Figure \ref{fig:parameterplot} evidences the  influence  of each parameter on the percentage of consumed calcium carbonate. The numerical derivatives of this percentage with respect to each parameter is plotted, providing insight into the sensitivity of the reaction to parameter variations. As expected, increasing either the reaction rate $\lambda$, corresponding to an acceleration time, or the diffusion coefficient $\overline{\sigma}$, leads to a higher percentage of calcium carbonate consumption. Meanwhile, the coefficient of the interaction with the environment term $\overline{\gamma}$ increases the amount of calcium consumed until it reaches the value $\overline{\gamma}=100$, beyond which the effect plateaus, suggesting a stabilization in the response of the system. The geometry of the system is then  influenced, concentrating the reaction of the surface of calcium carbonate.
 
\begin{table}[h!]
    \centering {\footnotesize
    \begin{tabular}{|c|c|c|c||c|c|c|c|}
        \hline
        $\lambda$ & $\overline{\sigma}$ & $\overline{\gamma}$ & $\kappa$ & $\lambda$ & $\overline{\sigma}$ & $\overline{\gamma}$ & $\kappa$ \\
        \hline
        50    & 0.1  & 100  & 0.0123 & 5000  & 0.1  & 100  & 0.2813 \\
        100   & 0.1  & 100  & 0.0220 & 5000  & 0.1  & 100  & 0.3093 \\
        500   & 0.1  & 100  & 0.0820 & 5000  & 0.2  & 100  & 0.3660 \\
        1000  & 0.1  & 100  & 0.1470 & 5000  & 0.4  & 100  & 0.4980 \\
        5000  & 0.1  & 1    & 0.2788 & 5000  & 0.6  & 100  & 0.5641 \\
        5000  & 0.1  & 10   & 0.2593 & 5000  & 0.8  & 100  & 0.6400 \\
        5000  & 0.1  & 100  & 0.2813 & 5000  & 1.0  & 100  & 0.6734 \\
        5000  & 0.1  & 100  & 0.3093 & 5000  & 1.5  & 100  & 0.6793 \\
        5000  & 0.1  & 1000 & 0.3083 & 5000  & 0.1  & 10000 & 0.2805 \\
        5000  & 0.1  & 15   & 0.2518 & 5000  & 0.1  & 1500 & 0.2854 \\
        5000  & 0.1  & 20   & 0.2714 & 5000  & 0.1  & 2000 & 0.2613 \\
        5000  & 0.1  & 250  & 0.2527 & 5000  & 0.1  & 30   & 0.2776 \\
        5000  & 0.1  & 3000 & 0.2540 & 5000  & 0.1  & 50   & 0.2735 \\
        5000  & 0.1  & 500  & 0.2907 & 5000  & 0.1  & 750  & 0.2696 \\
        10000 & 0.1  & 100  & 0.3100 & 50000 & 0.1  & 100  & 0.3614 \\
        50000 & 0.1  & 100  & 0.3619 & 100000 & 0.1  & 100  & 0.3816 \\
        \hline
    \end{tabular}}
    \caption{Sets of parameters and the percentage of calcium carbonate consumed. }
    \label{tab:parameterexploration}
\end{table}

\section{Conclusion}\label{Sec:conclusion}
The hybrid model we proposed combines the advantages of stochastic microscopic modeling and a continuous approach on the macroscale to obtain a more realistic description of the sulphation phenomenon. 
Our simulations showed realistic features of the corrosion process, including the environment-dependent movement of particle and the uneven progression of corrosion despite symmetric initial conditions.  These results are consistent with observed physical behaviors, suggesting the model’s potential to describe the underlying mechanisms of marble deterioration.
Future work will focus on analyzing the system in the limit of an infinite number of particles and investigating homogenization techniques to derive effective macroscopic descriptions of the corrosion process. 

It is worth noting that the hybrid modeling approach proposed here is not specific to the sulphation of marble -- many of the proposed modeling ingredients  can be used to describe other discrete-continuum  couplings as they arise, for instance, in the transport of charged particles through a polymer-based environment (like the morphology of an organic solar cell).  

\section*{Acknowledgments} 
N.J. and A.M. are involved in the Swedish Energy Agency’s project Solar
Electricity Research Centre (SOLVE) with grant number 52693-1, which offers a partial financial support. D.M, G.R. and S.U. carried the research project PON 2021(DM 1061, DM 1062) ``Deterministic and stochastic mathematical modelling and data analysis within the study for the indoor and outdoor impact of the climate and environmental changes for the degradation of the Cultural Heritage" of the Università degli Studi di Milano and are  members of GNAMPA (Gruppo Nazionale per l’Analisi Matematica, la Probabilità e le loro Applicazioni) of the Italian Istituto Nazionale di Alta Matematica (INdAM). A.M and D.M. thanks INdAM for  partially support (CUP E53C22001930001). 

\printbibliography
\end{document}